\documentclass[a4paper,fleqn,usenatbib]{mnras}

\usepackage[T1]{fontenc}
\usepackage{ae,aecompl}

\usepackage{graphicx}	% Including figure files
\usepackage{amssymb}	% Extra maths symbols

\usepackage{threeparttablex} %%Usar "tablenotes"
\usepackage{epsfig}
\usepackage{lmodern}

\newcommand{\bb}{\hbox{\it B\/}}
\newcommand{\vv}{\hbox{\it V\/}}

\newcommand{\ii}{\hbox{\it I\/}}
\newcommand{\jj}{\hbox{\it J\/}}
\newcommand{\kk}{\hbox{\it K\/}}
\newcommand{\wes}{\hbox{\it W\/}}

\newcommand{\bmv}{\hbox{\bb--\vv\/}}
\newcommand{\bmi}{\hbox{\bb--\ii\/}}

\newcommand{\rh}{\hbox{$r_{\rm h}$}}
\newcommand{\rc}{\hbox{$r_{\rm c}$}}

\newcommand{\logp}{\hbox{$\log P$}}
\newcommand{\feh}{\hbox{[Fe/H]}}

\title[]{Variable Stars in Local Group Galaxies. I: Tracing the Early Chemical 
Enrichment and Radial Gradients in the Sculptor dSph with RR Lyrae Stars.}

\author[C. E. Mart{\'i}nez-V{\'a}zquez et al.]{
C. E. Mart{\'i}nez-V{\'a}zquez$^{1,2}$,\thanks{E-mail: cmartinez@iac.es (CEMV)}
M. Monelli $^{1,2}$,
G. Bono$^{3,4}$,
P. B. Stetson$^{5}$, 
\newauthor{
I. Ferraro$^{4}$,
E. J. Bernard$^{6}$,
C. Gallart$^{1,2}$,
G. Fiorentino$^{7}$,
G. Iannicola$^{4}$,}
\newauthor{ and
A. Udalski$^{8}$}
\\
$^{1}$ IAC- Instituto de Astrof\'isica de Canarias, Calle V\'ia Lactea s/n, E-38205 La Laguna, Tenerife, Spain\\ 
$^{2}$ Departmento de Astrof\'isica, Universidad de La Laguna, E-38200 La Laguna, Tenerife, Spain\\ 
$^{3}$ Department of Physics, Universit\`a di Roma Tor Vergata, via della Ricerca Scientifica 1, I-00133 Roma, Italy\\
$^{4}$ INAF-Osservatorio Astronomico di Roma, via Frascati 33, I-00040 Monte Porzio Catone, Italy\\
$^{5}$ NRC-Herzberg, Dominion Astrophysical Observatory, 5071 West Saanich Road, Victoria BC V9E 2E7, Canada\\
$^{6}$ Institute for Astronomy, University of Edinburgh, Royal Observatory, Blackford Hill, Edinburgh EH9 3HJ, UK\\
$^{7}$ INAF-Osservatorio Astronomico di Bologna, via Ranzani 1, I-40127 Bologna, Italy\\
$^{8}$ Warsaw University Observatory, Al. Ujazdowskie 4, 00-478 Warszawa, Poland
}

\date{Accepted 2015 August 26.  Received 2015 August 25; in original form 2015 July 9}

\pubyear{2015}

\begin{document}
\label{firstpage}
\pagerange{\pageref{firstpage}--\pageref{lastpage}}
\maketitle


\begin{abstract} 

We identified and characterized the largest (536) RR Lyrae (RRL) sample
in a Milky Way dSph satellite (Sculptor) based on optical photometry data collected 
over $\sim$24 years. 

The RRLs display a spread in \vv-magnitude ($\sim$0.35 mag) which 
appears larger than photometric errors and the horizontal branch (HB) luminosity evolution
of a mono-metallic population. Using several calibrations of two different 
reddening free and metal independent
Period-Wesenheit relations we provide a new distance estimate
$\mu$=19.62 mag ($\sigma_{\mu}$=0.04 mag) that agrees well with 
literature estimates. We constrained the metallicity 
distribution of the old population, using the $M_I$ Period-Luminosity relation,
and we found that it ranges from -2.3 to -1.5 dex. The current
estimate is narrower than suggested by low and intermediate
spectroscopy of RGBs ($\Delta$ \feh $\le$ 1.5).

We also investigated the HB morphology as a function of the galactocentric
distance. The HB in the innermost regions is dominated by red HB stars 
and by RRLs, consistent with a more metal-rich population, while in the 
outermost regions it is dominated by blue HB stars and RRLs typical of a 
metal-poor population. Our results suggest that fast chemical evolution occurred in
Sculptor, and that the radial gradients were in place at an early epoch.

\end{abstract}


%%%%%%%%%%%%%%%%%%%%%%%%%%%%%%%%%%%%%%%%%%%%%%%%%%

%%%%%%%%%%%%%%%%% BODY OF PAPER %%%%%%%%%%%%%%%%%%

\section{Introduction}\label{introduction}

Understanding the evolution of Local Group dwarf galaxies offers a
fundamental key to constrain the early formation and evolution of 
cosmic structures \citep[e.g.][]{Mayer2010, Madau2014}. Current ground-based and space facilities 
allow us to resolve their stellar content and to provide a wealth of 
observables to reconstruct their star formation history and chemical 
enrichment \citep{Gallart2005,Tolstoy2009}.  

RR Lyrae (RRL) stars are reliable distance indicators, since their individual distances 
can be estimated using either the different developments of the Baade-Wesselink 
method \citep{Storm2004}, the correlation between visual magnitude and Fe 
abundance \citep{Cacciari2003}, 
statistical parallaxes \citep{Dambis2014}, optical, near-infrared (NIR) and 
mid-infrared (MIR) Period-Luminosity-Metallicity (PLZ) 
\citep{Bono2001, Bono2003, Catelan2004} and the reddening free 
Period-Wesenheit-Metallicity (PWZ, \citealt{Braga2015,Coppola2015:sub}). 
The above diagnostics all have pros and cons, but during the last few years the 
use of accurate and precise NIR and MIR mean magnitudes has provided a new spin 
on the use of RRLs as distance indicators \citep{Madore2013,Klein2014,Neeley2015}.     
This evidence was soundly complemented by the recent estimates of the 
trigonometric parallaxes for five field RRLs using the FGS on board 
of HST \citep{Benedict2011}. Moreover, the use of the reddening free
magnitudes, the so-called Period-Wesenheit relation (PWR) is firmly supported 
by theory \citep{Marconi2015} and observations \citep{Braga2015}.  

RRL stars also play a fundamental role as stellar tracers. Theory and 
observations indicate that they are present in all stellar systems 
hosting an old ($>$ 10 Gyr) stellar population. They are low-mass 
($0.6 - 0.8$M\sun) central helium-burning stars in their  
Horizontal Branch (HB) evolutionary phase. This ancient population is
the fossil record of the early stages of galaxy evolution and provides 
firm constraints on the time-scale of their early formation and evolution.

In this context, Sculptor is an interesting laboratory, since it hosts
a conspicuous old stellar population, but its star formation is quite 
complex \citep{DaCosta1984, Majewski1999, Tolstoy2004,deBoer2012}. 
Moreover, Sculptor appears to be characterized by a complicated chemical 
enrichment history \citep{Smith1983, Majewski1999, Hurley-Keller1999, Starkenburg2013}. 
In particular, the presence of a large metallicity spread has been suggested 
in this galaxy. Detailed spectroscopic measurements \citep{Tolstoy2004,Walker2007,
Walker2009b} showed not only that the more metal-rich red giant branch (RGB)
stars (\feh > --1.7) are more centrally concentrated than more metal-poor ones,
leading to a significant metallicity gradient, but also that they have different
kinematics. More importantly, the fact that red HB (RHB) stars are 
more centrally concentrated than blue HB (BHB) stars suggests that a spread 
in metallicity is also present in the old stellar component (HB stars),
which in turn suggests a relatively fast early chemical evolution. Since the
stars that are now RR Lyrae stars were formed during an even shorter time than
the global HB population, observing a gradient in their pulsational properties
would put even more stringent constraints on the time-scale of the chemical
enrichment in Sculptor \citep{Bernard2008}.

Since the discovery of Sculptor by \citet{Shapley1938}, a number of estimates
of its distance have been proposed in the literature. After the initial approximate 
value by \citet{Shapley1938} based on photographic data, the first
accurate measurement dates back to the work of \citet{Hodge1965} who
derived a true distance modulus of 19.7$\pm$0.15 mag  using three different methods 
(luminosity of the RGB, two W Virginis, and three RRL stars). \citet{Kunkel1977} provided 
distance estimates based on the average magnitude of 24
stars on the HB with colours covering the RR Lyrae gap
(0.15$<$(\bmv)$_0<$0.52); they found a true distance modulus 
of 19.47$\pm$0.10 mag, slightly lower than the previous one.

More recently, \citet{Pietrzynski2008} provided a distance value of
$\mu$=19.67$\pm$0.14 mag by applying different theoretical and empirical
calibrations of the PLZ relation for RRLs
in the NIR (\jj, \kk) magnitudes, consistent with the \vv-band data
of RRLs and also in very good agreement with the results obtained by
\citet{Rizzi2002} based on the tip of RGB (TRGB, 19.64$\pm$0.08 mag) and
HB stars (19.66$\pm$0.15) based on optical photometry.

The search for variable stars in Sculptor dates back 
to more than half a century ago \citep{Baade1939, Thackeray1950}. The 
first systematic search was performed by \citet{vanAgt1978}, who identified 
more than 600 candidate variable stars, but provided 
periods only for $\sim$10\% of the sample. 

The most complete RRL catalogue to date was provided by \citet{Kaluzny1995} 
based on the OGLE-I survey observations \citep{Udalski1992}. 
They investigated the central region of Sculptor (15$\arcmin \times$15$\arcmin$) 
and identified and characterized 226 RRLs. Their pulsation properties 
are consistent with metal-poor chemical composition (\feh$<$--1.7) and a 
spread in metallicity. The spectroscopic follow-up by \citet{Clementini2005} 
confirmed this preliminary evidence based on low resolution (LR) spectra of 107 RRLs. 
In particular, they found that the metallicity distribution peaks at \feh$\sim$ --1.8, 
but covers more than 1.5 dex (--2.4$\le$\feh$\le$--0.8).

We take advantage of a large sample of RRLs, including some new discoveries, to 
provide a revised distance estimate for Sculptor and to constrain the parameters of an 
early chemical enrichment. The current sample is a factor of two larger than previously
analysed ones (536 vs 216\footnote{10 out of the 226 RRLs published by \citet{Kaluzny1995}
turned out to be either double identifications or non-variables stars.}). 
Moreover, we provide pulsation period, mode and mean
magnitudes in three different photometric bands (see \S~\ref{photmetry}). In \S~\ref{distance} 
we describe the approach for the new distance determination, while in \S~\ref{metallicity} 
we investigate the spread in metallicity of the old stellar population using the \ii-band 
Period-Metallicity relation of RRLs.

%%%%%%%%%%%%%%%%%%%%%%%%%%%%%%%%%%%%%%%%%%%%%%%%%%%%%%%%%%%%%%%%%%%%%%%%%%%%%%%%
\section[]{Photometric data set and RR Lyrae identification} \label{photmetry}

We used 4,404 calibrated images from the photometric database collected by one of us
(PBS), covering an extended area around the Sculptor dSph. The time
baseline of these data is nearly 24 years (from October 1987
until August 2011). These data are individual CCD images from many
ground-based telescopes: CTIO 0.9 m+[TI, Tek2K], LCO 1.0 m+FORD2 (OGLE-I survey), MPI/ESO
2.2m+WFI, ESO NTT 3.6m+SUSI,  CTIO 4m+ Mosaic2, SOAR 4.1 m+SOI. The full
data-set covers an area of $\sim$6 sq. deg. centred on the Sculptor dSph
galaxy. However, only the central region ($\sim$3 sq. deg) could be
properly calibrated. The DAOPHOT/ALLFRAME package of programs
\citep{Stetson1987,Stetson1994} was used to obtain the instrumental
photometry of the stars. A detailed description of the observations and
data reduction will be given in a forthcoming paper.

We performed the variability search over the full data set. An updated
version of the Welch-Stetson variability index \citep{Welch1993} was used
to identify candidate variable stars on the basis of our multi-band
photometry. From the list of 663 variable star
candidates, we have found 611 actual variable stars and 23 possible
variable stars. Out of these real variables, 536 pulsating
stars are located in the Instability Strip (IS) of a very extended HB of the Sculptor dSph
galaxy, with periods between 0.2 and 1.2
days. Out of these, 82 RRLs are new identifications, and for 320 RRLs 
all their pulsational parameters (period, mean magnitude, and
amplitude) are derived for the first time. This work increases by over a factor of two the number of the known
RRL to date in the Sculptor dSph. Our sample includes 216 RRLs of \citet{Kaluzny1995}.

We derived pulsational properties for all the RRL stars
from our \bb\vv\ii-Johnson/Cousins photometry. The search for the period
was carried out using a simple string-length algorithm
\citep{Stetson1998b}. Then a robust least-squares fit of a Fourier series
to the data refined the periods. The intensity-averaged magnitudes and
amplitudes of the mono-periodic light curves were obtained by
fitting the light curves with a set of templates partly based on the set
of \citet{Layden1999} following the same method described in
\citet{Bernard2009}. Thus, through the period and light-curve shape, we
identified: {\em i)} 289 $RRab$, pulsating in the fundamental mode; 20
of which are suspected Blazhko, \citep{Blazhko1907}); {\em ii)} 197
$RRc$, pulsating in the first-overtone mode; and  {\em iii)}  50 
possible multi-mode $RRd$ stars, pulsating in both modes simultaneously, 
although the classification in some cases was uncertain due to their relatively 
noisy or (very) poor light curves. The mean (maximum) number of points in the
light-curves of the RRL stars are 83, 52, and 21 (115, 182 and 28) respectively
 in \bb, \vv, and \ii. 

%%%%%%%%%%%%%%%%%%%%%%%%%%%%%%%%%%%%%%%%%%%%%%%%%%%%%%%%%%%%%%%%%%%%%%%%%%%%%%%%
\section[]{RR Lyrae Instability Strip} \label{rrl}

%~~~~~~~~~~~~~~~~~~~~~~~~~~ FIG 1 ~~~~~~~~~~~~~~~~~~~~~~~~~~~

\begin{figure*}
\includegraphics[scale=0.70]{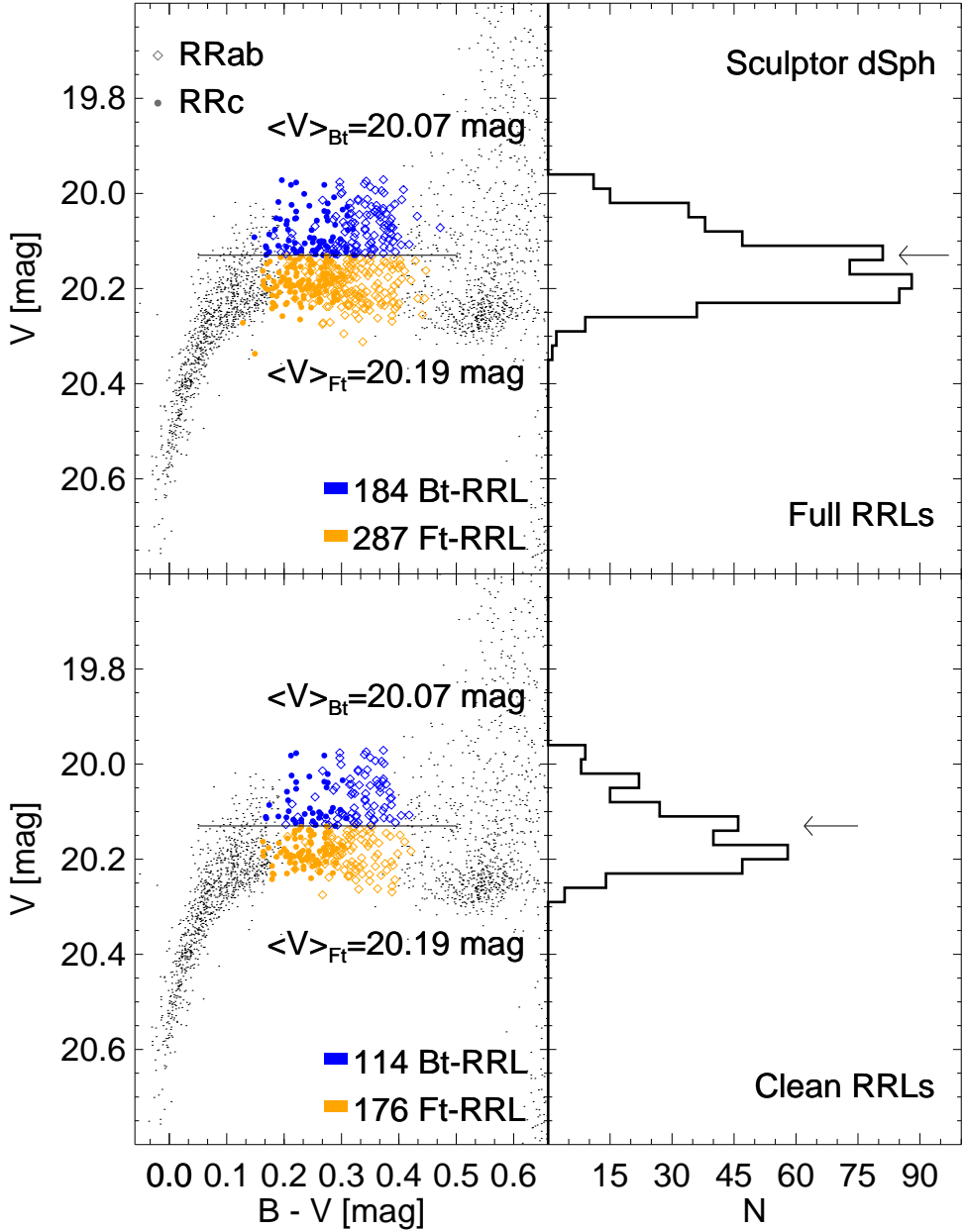}
 \caption{\textit{Top-left}. Optical (\bmv, \vv) CMD of Sculptor, 
zoom-in on HB stars. Blue and orange symbols display bright 
and faint RRLs of the \textit{full} sample. The horizontal line
shows the mean magnitude of the \textit{full} sample (\vv=20.13 mag) 
adopted to split the bright and faint samples.
Diamonds and circles display $RRab$ and $RRc$ 
variables. The number of RRLs in each subsample is labelled. 
\textit{Bottom-left}. Same as the top, but for the clean sample. 
\textit{Top-right}. \vv-band luminosity distribution of RRLs.  
The arrow marks the magnitude adopted to split bright and faint RRLs.
\textit{Bottom-right} Same as the top, but for the clean sample. 
For the sake of clarity, $RRd$ are not represented.}
\label{fig:fig1}
\end{figure*}

%~~~~~~~~~~~~~~~~~~~~~~~~~~~~~~~~~~~~~~~~~~~~~~~~~~~~~~~~~~~~~~~~~~~~~~~~~~~~~~~

%~~~~~~~~~~~~~~~~~~~~~~~~~~ FIG 2 ~~~~~~~~~~~~~~~~~~~~~~~~~~~

\begin{figure*}
\includegraphics[scale=0.70]{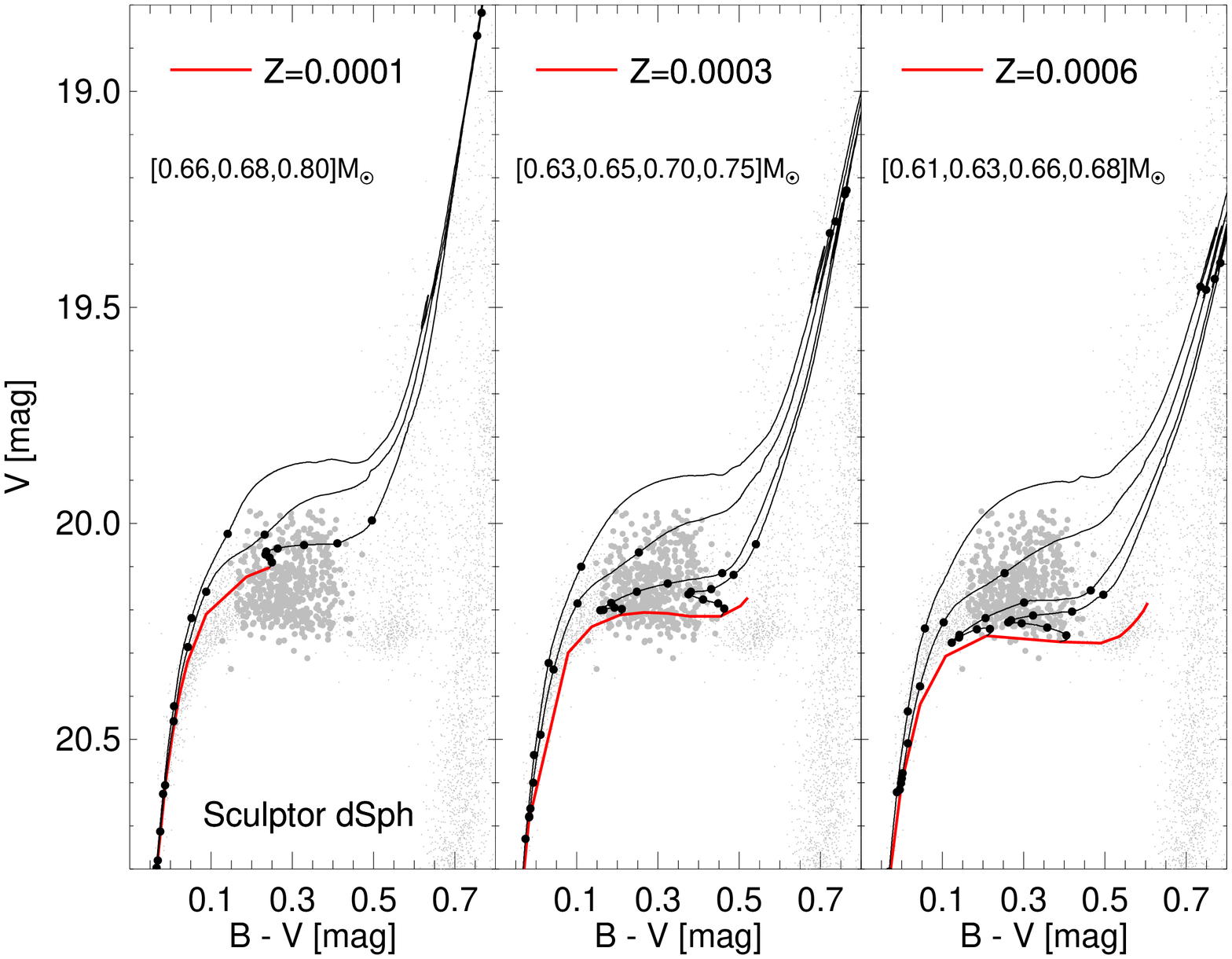}
 \caption{Optical (V,B-V) CMDs of Sculptor, zoom-in on HB stars. The grey circles 
display the RRL variable stars of the \textit{full sample} (same as top-left 
panel in Fig. \ref{fig:fig1}). The red lines show the zero age horizontal 
branch (ZAHB) from the BaSTI library \citep{Pietrinferni2004} for different
metallicities: Z=0.0001 (left), Z=0.0003 (middle), and Z=0.0006 (right).
The black lines display the corresponding evolutionary HB models with the
labelled masses, where the overplotted circles indicate time intervals
of 10 Myr. The stellar mass, at the typical colours of the RR Lyrae stars,
steadily increases when moving from the brightest to the faintest HB models.}\label{fig:tracks}
\end{figure*}

%~~~~~~~~~~~~~~~~~~~~~~~~~~~~~~~~~~~~~~~~~~~~~~~~~~~~~~~~~~~~~~~~~~~~~~~~~~~~~~~

The top left panel Fig.~\ref{fig:fig1} shows the (\bmv, \vv) colour-magnitude diagram (CMD) 
of Sculptor zoomed on the HB region. Data plotted in this CMD show a complex morphology 
with well defined BHB and RHB stars (black dots) and a large sample of RRLs. 
The latter group shows several interesting features.{\em i)} The RRL IS 
 is well populated over the entire colour range typical of $RRc$ and $RRab$ 
variables.{\em ii)} the spread in magnitude of RRLs is $\sim$0.35 mag, 
significantly larger than the typical uncertainty in the mean magnitudes 
($\sigma$=0.03 mag) and significantly larger than the expected from the evolution
of a mono-metallic population. The top-right panel of Fig.~\ref{fig:fig1} shows the visual magnitude 
distribution of the RRLs located within 2.5 $\sigma$ of the average. We ended up with a sample 
of 520 RRLs and neglected 16 outliers. Using the period-amplitude distribution 
and by visual inspection of the individual light curves we identified 
276 $RRab$ + 195 $RRc$ (defined by us as \textit{full sample}) and 49 candidate $RRd$ variables.
{\em iii)} The magnitude distribution of the RRLs is broad and possibly double peaked (see top-left panel). 

A plausible working hypothesis to account for the observed spread in 
magnitude of RRLs is that Sculptor RRLs cover a broad 
range in metallicity. Using classical luminosity-metallicity relations 
$M_{V}(RRL)=\alpha+\beta\cdot\feh$ by \citealt{Clementini2003} and
\citealt{Bono2003}, we found that a spread in \vv-magnitude of 0.35 mag implies 
a range in Fe abundance of the order of $\sim$0.6 dex. Note that this range 
has to be treated with caution, since the above relations are prone to systematic 
uncertainties, and in particular to evolutionary effects \citep{Bono2001, Bono2003}. 
   
In order to search for possible metallicity trends, we split the sample of 
RRab and RRc stars into two groups -- candidate RRd might have less accurate 
mean magnitudes and were thus neglected --. We have used the mean magnitude of 
the entire sample, $<$\vv$>$=20.13 ($\sigma$=0.09) mag, as the limit, shown 
by a solid line on the right-panel and the arrow on the right-panel of 
Fig.~\ref{fig:fig1}. We note that this selection is marginally affected by 
the curvature of the HB, in particular for the bluer stars, but that the results 
did not change significantly with a different criterion. We find 184 bright 
(Bt; blue symbols) and 287 faint (Ft; orange symbols) RRLs.
 
A more restrictive selection based on the quality of the phase coverage of the photometry 
over the entire pulsation cycle yielded similar results. This refined sample contains 290 RRLs that
we defined as the \textit{clean} sample. Among them we have 167 $RRab$ and 
123 $RRc$ variables. Data plotted in the bottom panels of Fig.~\ref{fig:fig1} 
indicate that the global properties are quite similar to the full sample, and indeed the  
mean \vv\ magnitude is $<$\vv$>$=20.14($\sigma$=0.07) mag. The same conclusion applies 
to the mean \vv\ magnitude of both Bt- and Ft-RRLs (see labelled values).  

In order to rule out the hypothesis that evolutionary effects can explain the 
observed luminosity range, Figure \ref{fig:tracks} shows a comparison with
stellar evolutionary tracks for different metallicities. The current predictions 
come from the BaSTI library \citep{Pietrinferni2004}, they have been computed 
assuming a scaled-solar chemical mixture (using the $\alpha$-enhanced HB 
models provide consistent results). They were plotted assuming the reddening law from 
\citet{Cardelli1989} and assuming a true distance modulus of $\mu$ = 19.62 mag 
(see \S~\ref{distance}) and a reddening of 0.018 mag.
The three panels present the case for Z=0.0001, 0.0003, 0.0006. In each 
panel the red thick line shows the Zero Age Horizontal Branch (ZAHB), while 
individual evolutionary tracks for selected labelled masses are presented by 
the thin black lines.
The black circles mark the points of each tracks with a time step of 10 Myr 
after the He ignition. Overall, the figure shows many interesting points: \\
{\em i)} The luminosity of HB stars mainly depends on the metallicity, 
therefore it is plausible to assume that the faint sample previously selected 
includes the most metal-rich RRL stars, while the bright sample should 
include the more metal-poor stars. However, off-ZAHB evolution could move more 
metal-rich stars into the bright sample. \\
{\em ii)} The comparison with tracks with Z=0.0006, in the right panel, 
suggests that this is a fair upper limit to the RRL metallicity. Nevertheless, 
the ZAHB is not able to well reproduce the RHB, suggesting that even more 
metal-rich stars populate this feature of the CMD.\\
{\em iii)} The left panel (Z=0.0001) shows that very metal-poor stars are
expected to start core Helium-burning on the blue side of the IS, except possibly 
the most massive ones. Nevertheless, the stars evolving off the ZAHB do cross
the IS, although this occurs at relatively bright 
luminosity where only a few very bright RRL stars are observed. Moreover,
the spacing between black points indicates that the crossing of the IS is an 
extremely rapid phase, thus very difficult to observe.
This is expected to be even more effective for extremely metal poor stars 
(Z<0.0001). Therefore, although possible, it seems unlikely that many of the
Sculptor RRL stars are as metal-poor as the most metal-poor stars detected
in the RGB \citep{Starkenburg2013}. \\
{\em iv)} While the tracks corresponding to each metallicity cover a significant
range of luminosity, it is clear that the luminosity spread associated to the
evolutionary effects of a mono-metallic population is not sufficient to
explain the observed magnitude range of RRL stars. This strongly 
supports our hypothesis that the RRLs in Sculptor cover a wide range of 
metallicities. \\

%~~~~~~~~~~~~~~~~~~~~~~~~~~ FIG 3 ~~~~~~~~~~~~~~~~~~~~~~~~~~~

\begin{figure*}
\includegraphics[scale=0.70]{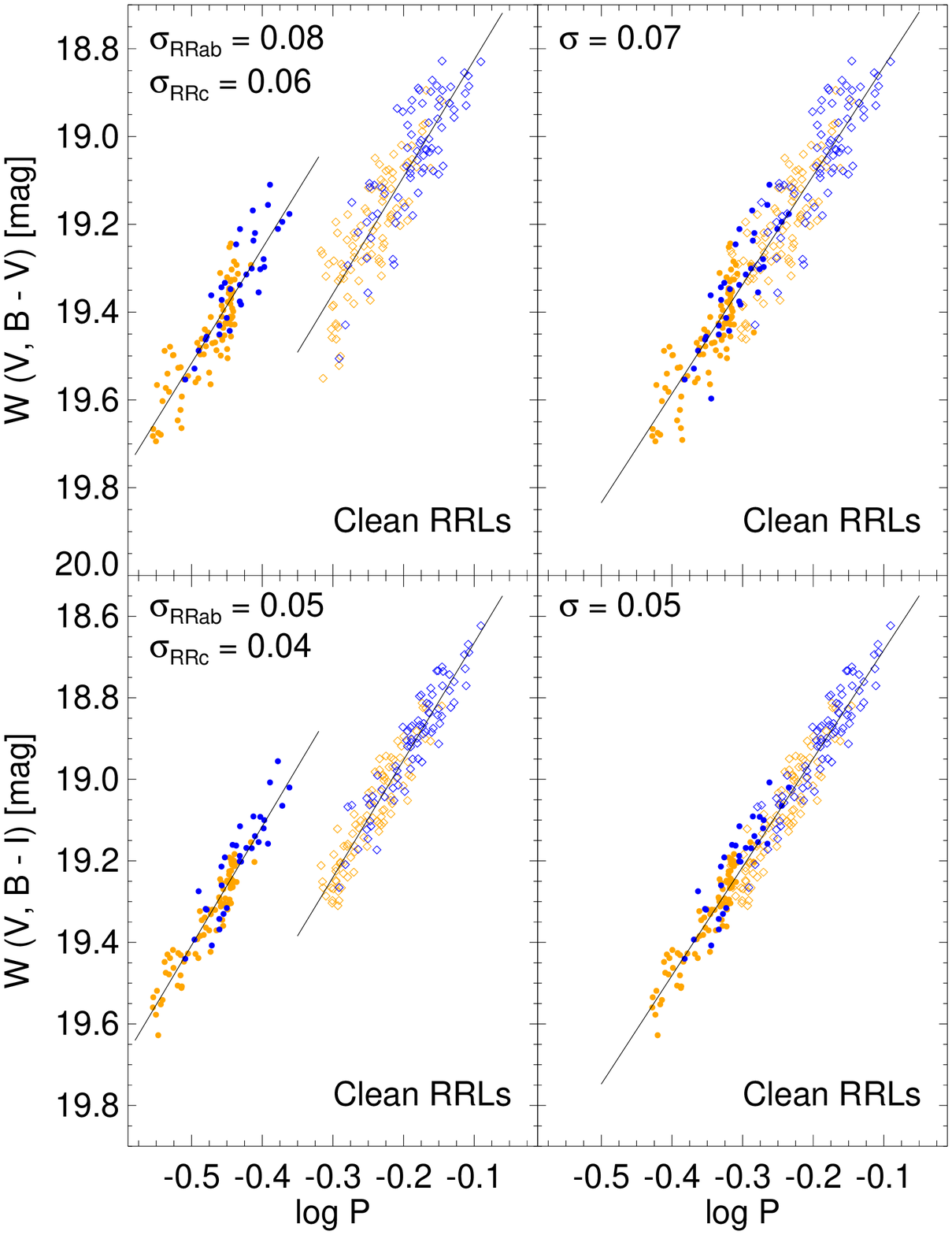}
\caption{\textit{Top-left}. Period-Wesenheit (\vv,\bmv) relation 
for $RRab$ and $RRc$ RRLs of the \textit{clean} sample. Symbols are 
the same as Fig.~\ref{fig:fig1}. Black lines display the least squares 
fits of the two subsamples. The \textit{sigma} of the fits are labelled.
\textit{Top-right}. Same as the left, but for 
the global ($RRab$+$RRc$) sample fundamentalized. \textit{Bottom-left}. Same as 
the top left, but for the PW (\vv, \bmi) relation. 
\textit{Bottom-right}. Same as the bottom left, but for the global
RRL sample fundamentalized.}\label{fig:wese}
\end{figure*}

%~~~~~~~~~~~~~~~~~~~~~~~~~~~~~~~~~~~~~~~~~~~~~~~
\begin{table}
\begin{scriptsize}
\centering
 \caption{Parameters of the theoretical fit, \wes(\vv,\bmi) = $\alpha$ + $\beta$ $\cdot$ \logp [mag], performed 
 following the procedure in \citet{Marconi2015} for metal-independent Wesenheit magnitudes.} 
 \label{tab:coef_th}
\hspace{+0.5cm}
  \begin{tabular}{l c c c } 
 \hline
\textbf{TH  \wes(\vv,\bmi)}    & $\alpha$ & $\beta$  & $\sigma$ \\  
 \hline
  \bf{$RRc$}    & -1.78$\pm$0.06      & -3.12$\pm$0.13  & 0.06 \\  
  \bf{$RRab$}  & -1.134$\pm$0.008  &-2.60$\pm$0.04   & 0.07 \\
   \bf{Global}    & -1.134$\pm$0.009  & -2.48$\pm$0.04  & 0.08 \\
\hline
\end{tabular}
\end{scriptsize}
\end{table}

\begin{table*}
\begin{scriptsize}
\centering
 \caption{Parameters of the empirical fit, \wes = $\alpha$ + $\beta$ $\cdot$ \logp [mag], performed on each group of RRL stars.} 
 \label{tab:coef_emp}
\hspace{-0.5cm}
  \begin{tabular}{l c c c c c c} 
 \hline
\textbf{EM}       & \multicolumn{3}{c}{CLEAN } & \multicolumn{3}{c}{FULL }  \\ 
 \hline
  \bf{$RRc$}          & $\alpha$  & $\beta$  & $\sigma$ & $\alpha$  & $\beta$ & $\sigma$  \\
 \hline
 \wes(\vv,\bmv)       & 18.21$\pm$0.06 & -2.61$\pm$0.13 & 0.06  & 18.26$\pm$0.05 & -2.48$\pm$0.10 &  0.06\\
\wes(\vv,\bmi)         & 17.95$\pm$0.05 & -2.92$\pm$0.10 & 0.04  & 18.05$\pm$0.04 & -2.65$\pm$0.08 &  0.05\\
\hline	
  \bf{$RRab$}            & $\alpha$  & $\beta$ & $\sigma$ & $\alpha$  & $\beta$  & $\sigma$  \\
 \hline
 \wes(\vv,\bmv)        & 18.56$\pm$0.03 & -2.66$\pm$0.12 & 0.08  & 18.54$\pm$0.02 & -2.72$\pm$0.10 &  0.09\\
 \wes(\vv,\bmi)        & 18.38$\pm$0.02 & -2.88$\pm$0.07 & 0.05  & 18.38$\pm$0.02 & -2.83$\pm$0.07 &  0.06\\
\hline
  \bf{Global} & $\alpha$   & $\beta$  & $\sigma$ & $\alpha$  & $\beta$  & $\sigma$  \\
 \hline
 \wes(\vv,\bmv)        & 18.59$\pm$0.02 & -2.48$\pm$0.05 & 0.07  & 18.58$\pm$0.01 & -2.50$\pm$0.05 &  0.08\\
 \wes(\vv,\bmi)        & 17.42$\pm$0.01 & -2.66$\pm$0.04 & 0.05  & 18.44$\pm$0.01 & -2.54$\pm$0.03 &  0.06\\
 \hline
\end{tabular}
\end{scriptsize}
\end{table*}
%%%%%%%%%%%%%%%%%%%%%%%%%%%%%%%%%%%%%%%%%%%%%%%%%%%%%%%%%%%%%%%%%%%%%%%%%%%%%%%%
\section[]{Distance determination} \label{distance}

\begin{table*}
\begin{scriptsize}
 \begin{minipage}{140mm}
  \caption{Distance moduli (in mag.) based on the current sample of RRL.}  
   \label{tab:dist}
  \hspace{-2.5cm}
  \begin{tabular}{ l c c c c c c} 
\hline
 & \multicolumn{3}{c}{CLEAN } & \multicolumn{3}{c}{FULL }  \\ 
\hline
\bf{$RRab$} & $\mu$ (\vv, \bmv) [165] & $\mu$ (\vv, \bmi) [164] & $<\mu>$  & $\mu$ (\vv, \bmv) [261] & $\mu$ (\vv, \bmi) [257] & $<\mu>$  \\
\hline
TH & 19.64$\pm$0.02(0.08) & 19.57$\pm$0.01(0.05) & & 19.63$\pm$0.02(0.09) & 19.57$\pm$0.01(0.06) &  \\
SE     & 19.64$\pm$0.02(0.08) & 19.51$\pm$0.01(0.05) & 19.59(0.05) & 19.61$\pm$0.02(0.09) & 19.52$\pm$0.01(0.06) & 19.58(0.04) \\
EM  ($\alpha=$[-1.03$\pm$0.10; -1.04$\pm$0.10]) & 19.59$\pm$0.02(0.08) & --- &   & 19.58$\pm$0.02(0.09)  & --- &  \\						   
\hline
\bf{$RRc$}            & $\mu$ (\vv, \bmv) [116] & $\mu$ (\vv, \bmi) [114] & $<\mu>$  & $\mu$ (\vv, \bmv) [178] & $\mu$ (\vv, \bmi) [177] & $<\mu>$ \\
\hline
TH       & 19.70$\pm$0.02(0.06) & 19.63$\pm$0.01(0.04) & & 19.69$\pm$0.02(0.06) & 19.62$\pm$0.01(0.05) &  \\
SE     & 19.77$\pm$0.02(0.06) & 19.73$\pm$0.01(0.04) & 19.67(0.10) & 19.82$\pm$0.02(0.08) & 19.84$\pm$0.01(0.05) & 19.70(0.14)  \\
EM   ($\alpha=$[-1.31$\pm$0.17; -1.24$\pm$0.17]) & 19.52$\pm$0.02(0.06) & --- &  & 19.51$\pm$0.03(0.06) & --- &  \\						   
\hline	
\bf{Global} & $\mu$ (\vv, \bmv) [279] & $\mu$ (\vv, \bmi) [277] & $<\mu>$ & $\mu$ (\vv, \bmv) [437] & $\mu$ (\vv, \bmi) [429] & $<\mu>$ \\
\hline
TH & 19.66$\pm$0.02(0.07)  & 19.60$\pm$0.01(0.05) & & 19.65$\pm$0.02(0.08) & 19.60$\pm$0.01(0.06) &  \\
SE & 19.66$\pm$0.02(0.07) & 19.55$\pm$0.01(0.05) & 19.61(0.05) & 19.65$\pm$0.02(0.08) & 19.58$\pm$0.01(0.06) & 19.61(0.04)\\
EM ($\alpha=$[-0.98$\pm$0.09; -0.98$\pm$0.09]) & 19.57$\pm$0.02(0.07) & --- &  & 19.56$\pm$0.02(0.08) & --- & \\						   
\hline
$<\mu>_{ADOPTED}$ & & & 19.62(0.04) & & & 19.63(0.06) \\ 
\hline
\end{tabular}
\end{minipage}
\end{scriptsize}
\end{table*}

The evidence that the RRLs in Sculptor show a spread in chemical composition 
is a thorny problem for distance determinations. 
The diagnostics adopted to estimate individual RRL distances depend 
on the Fe content (see \S~1). However, recent theoretical 
\citep{Marconi2015} and empirical \citep{Braga2015,Coppola2015:sub}
evidence indicates that the Period-Wesenheit relation (PWR) in the (\vv, \bmv) and the 
(\vv, \bmi) bands are minimally affected by metallicity. 
The above Wesenheit magnitudes are defined as 
\wes(\vv, \bmv)=\vv-3.06(\bb--\vv) and \wes(\vv, \bmi)=\vv--1.34(\bmi), where 
the coefficients were fixed according to the extinction law provided by \citet{Cardelli1989}
calculated for the effective mean wavelengths
of Landolt's \citep{Landolt1992} filter/photocathode combination 
for stars of spectral type A0III-F5III.

In their investigation \citet{Marconi2015} provided a detailed analysis
of optical, optical-NIR and NIR Period-Wesenheit-Metallicity (PWZ)
relations of the form  \wes(X, X--Y) = $\alpha$ + $\beta$ $\cdot$ \logp +
$\gamma$ $\cdot$ \feh, where X and Y are magnitudes. They found that the coefficient
of the metallicity term was smaller than $\sim$ 0.05 dex for the \wes(\vv,\bmv)
and the \wes(\vv, \bmi) PWR (see their Tables~7 and ~8), indicating that
the metallicity dependence of the \vv-band is counterbalanced by the
convolution between the reddening vector and (\bmv)-(\bmi) colors.

The top-left panel of Fig.~\ref{fig:wese} shows the PWR(\vv, \bmv) of the
\textit{clean} RRL sample. As expected $RRc$ and $RRab$ variables display 
different PWRs. The vanishing dependence of the quoted PWRs 
on metallicity is further supported by the small dispersion in apparent 
Wesenheit magnitude of RRLs for each period.  
We performed a linear least squares fit to the three different subsamples
and also applied a 3-$\sigma$ clip, and the final linear regressions are 
plotted as black lines in the top panels of Fig.~\ref{fig:wese}. The coefficients of 
the empirical PWRs are listed in Table~\ref{tab:coef_emp}. To constrain the 
impact of using the \textit{clean} sample on the empirical 
PWRs, the least squares fits were also performed using the 
\textit{full} RRL sample. Data listed in Table~\ref{tab:coef_emp} clearly indicate that 
the coefficients of the PWRs attain similar values within the 
uncertainties. 

We can see how the Bt-RRLs and Ft-RRLs follow the same relation. 
Moreover, Bt-RRLs seem to be the natural extension in the 
long-period regime of Ft-RRLs. The large sample of RRLs variables 
allows us to estimate the distance to Sculptor using three 
different PWRs for $RRc$, $RRab$, and the global sample ($RRc$+$RRab$, 
right panel of Fig.~\ref{fig:wese}). In deriving the distances of the global sample the 
periods of the RRc variables are fundamentalized, i.e.,
$\log P_F=\log P_{FO}+0.127$ \citep{Bono2001, Marconi2015}.

The \textit{bottom panels} of Fig.~\ref{fig:wese} display the PWR for RRLs in Sculptor, 
but in the \vv, \bmi\ bands (see Table~\ref{tab:coef_emp}). 
The standard deviations of the three different 
PWRs are slightly smaller than in \vv, \bb-\vv. The difference is 
mainly due to the fact that \bmi\ is a better temperature indicator 
and the PWR closely mimics a Period-Luminosity-colour relation 
(Bono et al. in preparation).

The distance estimates to Sculptor were derived using the three different 
subsamples  ($RRab$, $RRc$, and global) and both the \textit{clean, 290} 
and the \textit{full, 471} RRL stars. In order to quantify the effect of 
both the zero-point ($\alpha$) and the slope ($\beta$) 
of the relation on the distances, we adopted three different approaches: 
{\em i)} theoretical -- TH-- in which we adopted predicted PWRs \citep{Marconi2015};  
{\em ii)} semi-empirical -- SE-- in which we adopted the observed slopes
listed in Table~\ref{tab:coef_th} and the theoretical zero-points; 
{\em iii)} empirical -- EM -- in which we adopted the empirical slopes and where the 
zero-points are based on the field RRLs with HST trigonometric parallaxes 
\citep{Benedict2011}. In particular, we adopted RR Lyr itself to calibrate the 
$RRab$ and the global sample, while the $RRc$ were calibrated using RZ Cep.      
The mean \bb\ and \vv\ magnitudes provided by 
\citet{Coppola2015:sub} and the individual distance moduli provided by 
\citet{Braga2015} and \citealt{Neeley2015} have been used.

Using the empirical slopes and the above distances we estimated the   
zero-points for $RRab$, global and $RRc$ PW relations 
(see Table~\ref{tab:dist}). The former values for $RRab$ and global 
relations agree quite well with predicted ones (--1.08$\pm$0.01, --1.07$\pm$0.01). 
However, in the case of the $RRc$, the new zero-points are $\sim$30\% smaller than 
predicted (--1.56$\pm$0.04).

Table~\ref{tab:dist} gives the distance moduli based on the \textit{clean} and \textit{full} 
RRL samples interpreted with the three different PWRs ($RRab$, $RRc$, global) 
and the three different calibrations (TH, SE, EM). The number of RRLs 
used to estimate the distance after the 3$\sigma$ clipping is given in
square brackets. Table~\ref{tab:dist} gives true distance moduli (mean values 
of $\mu$ (\vv, \bmv) and $\mu$ (\vv, \bmi) over the entire sample. For each fixed sample 
and approach, the estimated distance moduli agree with each other within 1$\sigma$ and 
further support the evidence that the two \textit{clean} and \textit{full} samples give very 
similar distances. Since there is no reason {\it a priori} to prefer one of the five 
different estimates, for each sample we estimated the mean value ---denoted by $<\mu>$--- 
that is given with the relative standard deviation ($\sigma_{\mu}$). 
The distance determinations listed in Table~\ref{tab:dist} agree quite well with similar 
distances available in the literature.  In particular, they agree with the 
distance obtained by \citet{Pietrzynski2008} using RRLs [19.67$\pm$0.02(0.12)]. 
The difference is indeed smaller than 0.4$\sigma$. 
The true distance modulus to Sculptor was estimated as the mean among the mean 
distances of $RRc$, $RRab$ and global sample. The mean distance based on the global 
sample is treated as an independent determination, because the $RRc$, once 
fundamentalized, become ``pseudo'' fundamental pulsators \citep{Braga2015}. 
We found $<\mu>_{clean}$=19.62, $\sigma_{\mu}$=0.04 and $<\mu>_{full}$=19.63, 
$\sigma_{\mu}$=0.06 mag, respectively. We adopted the former one, for the
reasons mentioned above.

%%%%%%%%%%%%%%%%%%%%%%%%%%%%%%%%%%%%%%%%%%%%%%%%%%%%%%%%%%%%%%%%%%%%%%%%%%%%%%%%
\section[]{Metallicity of the RRL stars}\label{metallicity}

\subsection{Metallicity range}

%~~~~~~~~~~~~~~~~~~~~~~~~~~ FIG 4 ~~~~~~~~~~~~~~~~~~~~~~~~~~~

\begin{figure}
\hspace{-0.5cm}
\includegraphics[scale=0.53]{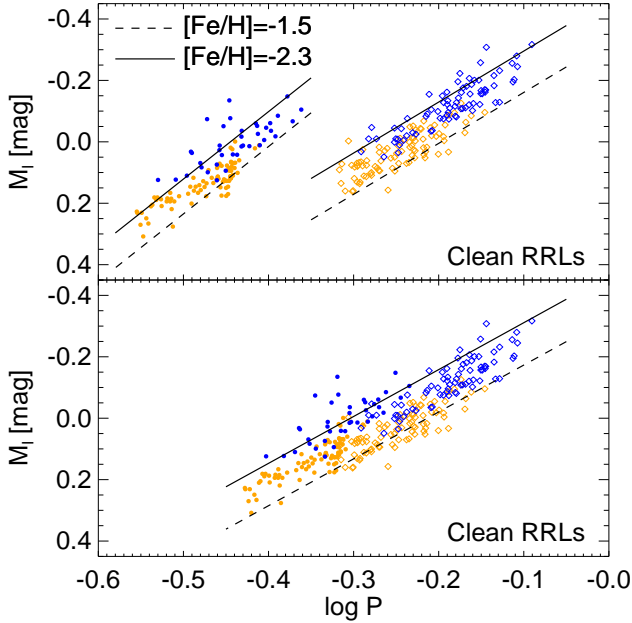}
\caption{\textit{Top}. Absolute $M_{\ii}$ Period-Luminosity relation for 
$RRab$ and $RRc$ variables of the clean sample. Symbols are the same as in 
Fig.~\ref{fig:fig1}. The dashed and solid black lines show predicted PLR 
for two different Fe abundances (\feh=-1.5, -2.3; \citep{Marconi2015}). 
\textit{Bottom}. Same as the top, but for the global sample.}\label{fig:pli}
\end{figure}

%~~~~~~~~~~~~~~~~~~~~~~~~~~~~~~~~~~~~~~~~~~~~~~~~~~~~~~~~~~~~

We will now take advantage of the new accurate distance determination to Sculptor [19.62$\pm$0.01(0.07)] and 
the dependence of the \ii-band Period-Luminosity (PLI) relation on metallicity 
\citep{Marconi2015} to constrain the metallicity distribution of the 
\textit{clean} RRL sample. The top-left panel of Fig.~\ref{fig:pli} shows the comparison 
between RRLs and predicted PLI relations for two different metal abundances  
\feh=--2.3 (solid line)  and \feh=--1.5 (dashed line). The majority of both $RRc$ 
and $RRab$ are bracketed by the quoted predictions. This evidence is further 
supported by the global sample (bottom-left panel). Moreover, the Bt-RRL
and Ft-RRL subsample are located closer to the metal-poor and metal-rich relations,
respectively. We performed the same test using the \textit{full} sample and 
found that the metallicity ranges from $\sim$ --2.4 to $\sim$ --1.4 dex.   

The spread in metallicity of the RRLs that we have measured using this method -- 0.8 to 1.0
dex -- is in excellent agreement with that obtained by \citet{Clementini2005} from spectroscopic measurements of RRL stars, but significantly smaller than the estimates based on spectroscopic metallicities of RGB stars. Measurements 
from LR spectra of 107 RRLs ($\Delta$S parameter) indicate a mean 
Fe abundance of --1.83$\pm$0.03 ($\sigma$=0.26) with the metallicity ranging 
from $\sim$ --2.40 to --0.85 dex \citep{Clementini2005}. However, they found only one
metal-rich (\feh=$\sim$ --0.85) RRL star while the bulk covers in a metallicity range
from $\sim$ --2.4 to --1.4 dex. Therefore, our estimates are in agreement with the bulk of 
the RRLs included in \citet{Clementini2005}. More detailed spectroscopic 
measurements from medium resolution spectra for more than 300 RGBs revealed 
the occurrence of a metal-poor stellar component with Fe abundance ranging 
from --2.8 to --1.7 and of a metal-rich component with metallicity ranging from --1.7 to 
--0.9 \citep{Tolstoy2004,Coleman2005}. The empirical scenario was 
recently fortified by the extensive spectroscopic investigation based on medium resolution
 spectra of almost 400 RGBs by \citet{Kirby2009}. They found a mean 
metallicity of --1.58 ($\sigma$=0.41) with the Fe content ranging from 
$\sim$ --3 to $\sim$ --1 dex. Moreover, they also found evidence of a linear 
metallicity gradient.

However, we cannot exclude the possible occurrence of small samples 
of more metal-poor or metal-rich old stars, because these stellar components might 
produce HB stars either too hot or too cool to inhabit the RRL IS.  This particularly means that 
RRLs are not sensitive to the possible occurrence of an extremely metal-poor 
component \citep[][see also Fig.~\ref{fig:tracks}]{Kirby2009,Starkenburg2013,Simon2015}. 
However, the lack of High Amplitude Short Period fundamental RRLs
\citep[HASP; Log P $\le$ -0.35, A$_V \ga$ 1,][]{Fiorentino2015} suggests that
Sculptor does not host a significant metal-rich (\feh$\ge$-1.3) old stellar
population (Mart\'{i}nez-V\'azquez et al. in preparation).

\subsection[]{Radial properties}\label{radialdep}

%~~~~~~~~~~~~~~~~~~~~~~~~~~ FIG 5 ~~~~~~~~~~~~~~~~~~~~~~~~~~~

\begin{figure}
\hspace{-1cm}
\includegraphics[scale=0.53]{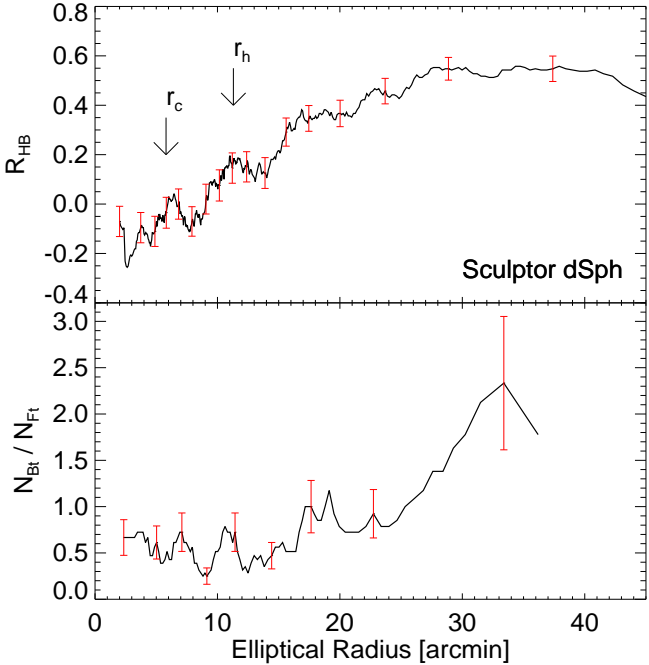}
\caption{\textit{Top}. The $R_{HB}$ parameter as a function of the the elliptical 
radius. The vertical arrows mark the core (\rc=5.8 arcmin, \citealt{Mateo1998}) 
and half-light radius (\rh=11.3 arcmin, \citealt{Irwin1995}). 
The vertical red lines display the errors on $R_{HB}$ based on Monte-Carlo simulations.   
The \textit{Bottom}. Same as the top, but for the ratio between 
bright and faint RRLs in the \textit{full} sample.}\label{fig:HBR}
\end{figure}

%~~~~~~~~~~~~~~~~~~~~~~~~~~~~~~~~~~~~~~~~~~~~~~~~~~~~~~~~~~~~

\citet{Tolstoy2004} found a radial gradient in the metallicity distribution of both 
RGB and HB stars, in the sense that the more metal-poor component is more spatially 
extended than the metal-rich one. Now, we will explore whether a radial gradient 
exists in the properties of the Sculptor RRL stars and whether it is associated to 
a metallicity gradient.

To avoid spurious fluctuations when moving from the innermost to the outermost 
regions we ranked the entire sample of RRL stars as a function of the radial distance. 
Then we calculated the running average by using a box starting with the first 200
more central objects in the list. We estimated the mean $R_{HB}$\footnote{We adopted the 
$R_{HB}$=(B--R)/(B+V+R) parameter \citep{Lee1990}, i.e., the ratio among stars 
that are located either inside (V=RRLs), or to the blue (B=BHBs), or to the red
(R=RHBs) of the RRL IS.}, radial distance and 
standard deviation in each successive sample of 200, removing five objects 
in the ranked list and adding five, until we reached the last 200 objects in the 
sample (651 RHBs, 536 RRLs, and 993 BHBs). We performed several tests 
changing both the number of objects included in the box and the step size 
and found that our conclusions are not affected. The data plotted in the
top panel of Fig. 5 show that the value of $R_{HB}$ parameter is steadily 
increasing over a substantial fraction of the body of the galaxy (r$<$25 arcmin), 
but flattens to RHB $\sim$0.55 in the outermost regions. 
The outermost regions display an almost constant HB morphology ($R_{HB}$ $\sim$0.55). 
The steady increase in the $R_{HB}$ suggests that we move from a HB mainly dominated 
by RRL and RHB stars ($R_{HB}$ $\sim$-0.1) inside the core radius (\rc) to a HB mainly 
dominated by BHB stars and RRLs ($R_{HB}$ $\sim$0.4) 
beyond the half-light radius (\rh).   

To further constrain the above intriguing evidence we also tested the 
radial trend of the ratio between Bt- and Ft-RRLs (\textit{clean} sample),
following the running average procedure with a sample size of 50, replacing five objects 
in the ranked list until we reached the last 50 objects in the 
sample of RRLs. The data plotted in the bottom panel 
show that the above ratio is quite constant ($N_{Bt}/N_{Ft}\sim$0.5) inside the 
\rh, but it increases by a factor of three in the outermost regions 
(r$\sim$25 arcmin). We cannot reach firm conclusions regarding a flattening 
in the outskirts of the galaxy (r$\sim$30 arcmin) due to the limited sample of RRLs in 
these regions. The above findings therefore suggest that the metallicity gradient 
was already present $>$10 Gyr ago.

%%%%%%%%%%%%%%%%%%%%%%%%%%%%%%%%%%%%%%%%%%%%%%%%%%%%%%%%%%%%%%%%%%%%%%%%%%%%%%%%
\section[]{Conclusions}\label{conclusions}

We have performed detailed multi-band homogeneous optical photometry of the Sculptor 
dSph. We accumulated more than 4,400 images (proprietary plus archive images) covering 
a time interval of $\sim$ 24 years. The precision of the photometry and the accuracy of both 
the absolute and relative zero-points allowed us to provide new mean magnitudes for 
574 variable stars. Among them 536 have been identified as Sculptor RRLs: 82 are new
identification while for 219 candidates we provide for the first time the pulsation parameters.
The new RRL data set allowed us to investigate several interesting open problems. 

{\em i) True distance modulus}. The use of three photometric bands and of two 
diagnostics, the PWRs in \vv, \bmv, and in \vv, \bmi, that are 
reddening free and only minimally affected by metal content, allowed us to 
provide very accurate distances. The errors on individual distance modulus are 
never larger than 0.02$_{sys}$ and 0.09$_{ran}$.
To investigate possible systematics affecting the quoted diagnostics, 
we have adopted three independent calibrations, named empirical, 
semi-empirical, and theoretical. We ended up with a mean distance moduli of 
19.62 with $\sigma_{\mu} =$ 0.04 using the \textit{clean} (290) sample. 
All the estimates presented in this paper agree quite well with similar 
evaluations based on reliable distance indicators \citep{Pietrzynski2008}.

{\em ii) Metallicity range}. We used the \ii-band PLR of RRL to constrain the
 metallicity range covered by Sculptor RRLs. The \textit{clean} 
RRL subsample suggests that the Fe content of the old (t$>$10Gyr) stellar 
population ranges from --2.3 to --1.5 dex. This range is in excellent 
agreement with spectroscopic measurements of RRLs, but significantly 
narrower than the distribution of RGB metallicities (--3.0$\la$\feh$\la$--0.8 dex, 
\citealt{Kirby2009}).

{\em iii) HB morphology}. The use of homogeneous and accurate photometry 
covering a significant fraction of the body of the galaxy allowed us to 
constrain the HB parameter as a function of radius. We found that the HB morphology 
changes from --0.1 (dominated by BHB stars and RRL) to +0.55 
(dominated by RHB stars and RRL) when moving from the innermost to the outermost 
galactic regions, as expected from the presence of a younger and/or more metal-rich stellar 
population in the center of the galaxy \citep[e.g.][]{deBoer2012}.

{\em iv) RRL luminosity distribution}. We split the RRL sample into 
Bt-RRL (\vv $\le$ 20.13 mag) and Ft-RRL (\vv $>$ 20.13 mag) sub-samples. 
The large sample of RRLs allowed us to constrain the radial gradient 
of the Bt/Ft RRL ratio. We found that this ratio is almost 
constant in the innermost regions while it shows a steady increase for 
radial distance ranging from \rh to 25 arcmin. In passing we note that this is, 
to our knowledge, the first clear evidence in which HB stars and RRLs display 
similar radial trends.  Analogous empirical evidence was found in the Tucana dSph 
\citep{Bernard2008}. However, a dichotomous population of RRLs was clearly identified 
there.

This work shows that RRLs provide significant constraints on the chemical enrichment
history of its host galaxy. In particular, the combination of a large sample -- one of the largest RRL 
samples in a dSph galaxy to date \citep{Bersier2002, Monelli2012}  -- with high quality, multi-band photometry 
allowed us to determine the metallicity distribution of the old stellar component, and revealed the presence of a significant metallicity gradient among the oldest stars ($>$10 Gyr).
Our results are in agreement with the star formation history of Sculptor recently obtained 
by \citet{deBoer2012}, the synthetic HB modelling of \citet{Salaris2013}, as well as the various 
works based on spectroscopic data \citep[see, e.g. ][]{Tolstoy2004, Battaglia2008b, Kirby2009}.
In particular, we find that Sculptor underwent a very efficient chemical enrichment: the metallicity range 
of RRLs shows an increase in the Fe content by almost one dex in less than 2--3 Gyr. The metallicity 
gradient suggests that star formation lasted until more recently in the inner regions, therefore further 
enriching the interstellar medium. However, we cannot exclude the possibility of a merger at an early 
epoch, or tidally-induced heating of an initially disk-like structure into a spheroid 
\citep{Mayer2001,Mayer2006} as the origin of the gradient.

\section*{Acknowledgments}
We are very grateful to our referee, Mario Mateo, for his useful
comments and suggestions that helped to improve the readability
and the content of this paper.
We thank S.~Cassisi for providing interesting and helpful 
discussions about the evolution of the stars in the horizontal branch.
CEMV thanks V.~F.~Braga for helping us with the theoretical 
relations, and is grateful to the Rome Observatory and the
Physics Department of the Tor Vergata University where most 
of this work has been realized.
This research has been supported by the Spanish Ministry of Economy and 
Competitiveness (MINECO) under the grant (project reference AYA2014-56795).
The OGLE project has received funding from the National Science Centre,
Poland, grant MAESTRO 2014/14/A/ST9/00121 to AU.
GF has been supported by the Futuro in Ricerca 2013 (grant RBFR13J716).

%%%%%%%%%%%%%%%%%%%%%%%%%%%%%%%%%%%%%%%%%%%%%%%%%%



\bibliographystyle{mnras}
%\bibliography{cmartinez_bibtex}

 \newcommand{\noop}[1]{}

%%%%%%%%%%%%%%%%%%%%%%%%%%%%%%%%%%%%%%%%%%%%%%%%%%

% Don't change these lines
\bsp	% typesetting comment
\label{lastpage}
\end{document}